\documentclass[prl,twocolumn,showpacs,amsmath,amssymb,superscriptaddress]{revtex4-1}

\usepackage{graphicx}
\usepackage{amsmath}
\usepackage{amssymb}
\usepackage{array}
\usepackage{SIunits}
\usepackage{color}
\usepackage{dcolumn}

\def\ket0{$\left|0\right>$}
\def\ket1{$\left|1\right>$}
\def\ca40{$^{40}\mathrm{Ca}^+$}
\def\T2{$\mathrm{T_2$}}
\def\Pr3{$\mathrm{Pr^{3+}}$}
\def\ket#1{$\left|#1\right>$}
\def\mket#1{\left|#1\right>}

\def\eq#1#2{\begin{equation}\label{Eq:#1}#2\end{equation}}

\def\fref#1{\ref{Fig:#1}}

\begin{document}
\newcommand*{\MAINZ}{Institut f\"ur Quantenphysik, Universit\"at Mainz, Staudingerweg 7, 55128 Mainz, Germany}
\newcommand*{\contrib}{These authors contributed equally to the work}
\newcommand*{\correspond}{contact: poschin@uni-mainz.de, www.quantenbit.de}

\title{Controlling fast transport of cold trapped ions}

\author{A.~Walther}\altaffiliation{\contrib}
\author{F.~Ziesel}\altaffiliation{\contrib}
\author{T.~Ruster}
\author{S.~T.~Dawkins}
\author{K.~Ott}
\author{M.~Hettrich}
\author{K.~Singer}
\author{F.~Schmidt-Kaler}
\author{U.~Poschinger}\altaffiliation{\correspond}

\affiliation{\MAINZ}

\date{\today}

\begin{abstract}	
We realize fast transport of ions in a segmented micro-structured Paul trap. The ion is shuttled over a distance of more than 10$^4$ times its groundstate wavefunction size during only 5~motional cycles of the trap (280~\micro m in 3.6~\micro s). Starting from a ground-state-cooled ion, we find an optimized transport such that the energy increase is as low as 0.10$\pm0.01$ motional quanta. In addition, we demonstrate that quantum information stored in a spin-motion entangled state is preserved throughout the transport. Shuttling operations are concatenated, as a proof-of-principle for the shuttling-based architecture to scalable ion trap quantum computing.
\end{abstract}


\maketitle
The field of quantum information began its experimental uprise with the proposal from Cirac and Zoller~\cite{CIRAC1995} in 1995, extended by the prospect of scalability of ion trap based systems~\cite{Blatt2008} via shuttling of qubits in multiplexed traps along the ideas pioneered by Wineland~\cite{KIELPINSKI2002}. Scalable information processing in a multiplexed ion trap can be accomplished by having fixed processing sites where logic operations are performed, and ion qubits will be moved in and out of these regions by shuttling operations. The duration of such shuttling has to be much faster than the relevant decoherence times~\cite{Monz2011}. Furthermore, it is desirable to reduce the total time consumption of all relevant operations, where shuttling will contribute a considerable amount~\cite{Steane2007}, and aim for the performance of the naturally fast solid state architectures~\cite{Mariantoni2011}. So far, ion shuttling in a multiplexed trap has been demonstrated together with additional sympathetic cooling~\cite{HOME2009}, and in the adiabatic regime, where the transient displacement of the ion is smaller than the size of the its wavepacket~\cite{Blakestad2009,Blakestad2011}. Transport of neutral atoms have also been performed using magnetic~\cite{Hansel2001} or optical~\cite{Kuhr2003} techniques.

Because quantum gate operations require ions close to the motional ground state and fast transport inherently creates motional excitation, the challenge is to develop transport protocols that guarantee sufficiency small energy transfer. In this work we demonstrate shuttling operations that are highly non-adiabatic while the final state of the ion is close to the motional groundstate. We also show that quantum information stored in both the motional and the spin degree of freedom is preserved through the shuttling.

\begin{figure}[ht]
\includegraphics[width=8.5cm,clip=true]{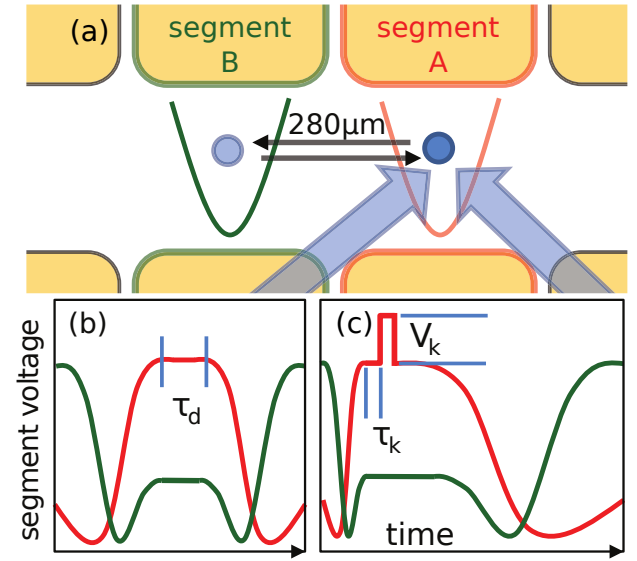} 
\caption{(color online) Schematic view of the trap and the voltage ramps (not to scale) that are used during the transport operations. In a), the segments used in the experiments are depicted together with the laser beams that are used for qubit interactions and for detection of the motional excitation. The size of each segment is 250~\micro m, the separation gap is 30~\micro m and the upper and lower rows are 500~\micro m apart. Part b) shows the pairwise energy-neutral transport ramp and c) the self-neutral transport (see text).}
\label{Fig:trap}
\end{figure}

During a shuttling operation, the ion motion in the harmonic trapping potential is excited when the acceleration is sufficiently strong. This motional excitation is a harmonic oscillation, characterized by a well defined phase, thus allowing it to be canceled out by proper management of the forces involved during or after the transport. We experimentally demonstrate two methods of canceling the acquired motional excitation. One method uses two shuttles, where the transport to the destination generates the same net momentum transfer as the transport back, but is applied 180$^\circ$ out of phase with respect to the secular oscillation of the ion (Fig.~\ref{Fig:trap}b). We refer to this as the \emph{pairwise energy-neutral} transport. For the second scheme, the \emph{self-neutral} transport we apply a sharp counter-``kick'' to the ion at the end of a single transport operation, stopping its motion (Fig.~\fref{trap}c). This case of single-sided transport allows even faster shuttling and can be sequentially repeated since it is energy self-neutral, making it the preferred building block for scalable quantum information protocols.

The experiments are carried out in a micro-structured segmented Paul trap~\cite{SCHULZ2008}, where the trapping voltage applied to each electrode segment is controlled by an FPGA-based arbitrary waveform generator that allows high resolution for both timing and voltage, while having low noise ($\lesssim 10 \mathrm{nV/\sqrt{Hz}}$ at the trap frequency). Further noise reduction at high frequencies is done with a $\mathrm{\Pi}$-type filter having a cut-off frequency of $300$~kHz. The axial trap frequency is $\nu_{\mathrm{ax}} = \omega_{\mathrm{ax}}/2\pi = 1.41$~MHz, while the radial trap frequencies are $\sim$~3~MHz. The transport is carried out by varying the voltages on the segments, such that the minimum of the trapping potential is moved towards the final location in time steps of 400~ns. In order to minimize parametric excitation, the confinement strength of the trap should remain as close as possible to its original value during transport. To assure this, the voltages required for creating a trap minimum with predefined frequency at a given location are calculated a priori. This relation is established using numerically obtained data for the electrostatic trap potentials~\cite{Singer2010a}, and has been experimentally verified~\cite{Huber2010}. The ion used for the experiments is \ca40, where an external magnetic field splits the $S_{1/2}$ ground state into two levels, $m_J = \pm 1/2$, henceforth referred to as \ket{\uparrow} and \ket{\downarrow}. Qubit rotations between these levels are mediated by stimulated Raman transitions~\cite{POSCHINGER2009,LEIBFRIED2003}. Each experimental run is started by i) Doppler cooling and followed by ii) optical pumping, leaving the ion in the \ket{\uparrow} state. iii) Resolved sideband cooling is then used to prepare the ion close to the motional ground state at a mean thermal phonon number of about $\bar{n}_{\mathrm{th}} \approx 0.1$ phonons, to which the motional excitation is then compared. iv) A transport operation is performed. v) For determining the motional state, stimulated Raman transitions between \ket{\uparrow} and \ket{\downarrow} are driven by a pair of off-resonant beams propagating at 90$^\circ$ with respect to each other. The effective wavevector of the beams is aligned along the trap axis, providing a coupling to the axial mode of vibration, characterized by a Lamb-Dicke factor of $\eta\approx 0.23$. vi) Spin read-out is performed by a shelving pulse  followed by the detection of state-dependent fluorescence~\cite{POSCHINGER2009}.

\begin{figure}[ht]
\includegraphics[width=8.5cm]{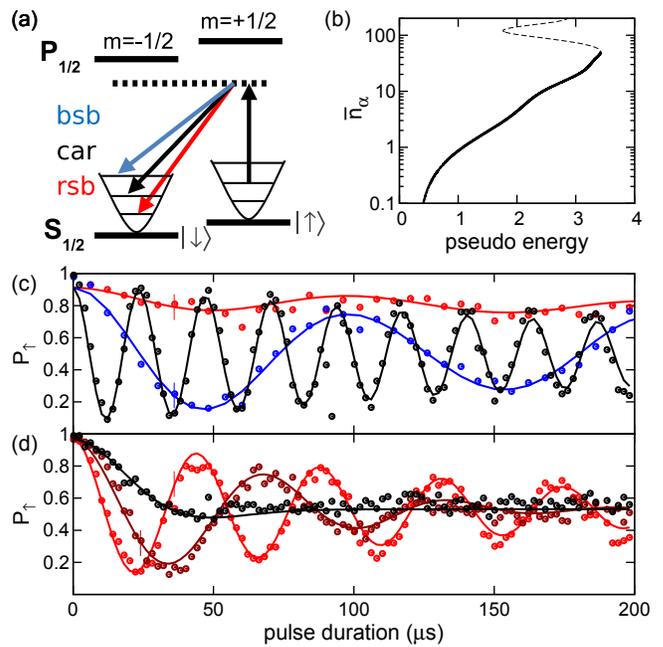}
\caption{(color online) Determination of the motional excitation. a) Relevant levels and transitions in the \ca40 system. For the determination of the phonon number we use two methods based on the excitation of sidebands. b) Phonon number as a function of the pseudo energy quantity (see text), indicating the bijective nature for $\bar{n}_{\alpha} \lesssim 50$. Rabi oscillations for the carrier (black), first blue (blue), first red (red) and second red (dark red) sideband, (c) for the case of $\bar{n}_{\alpha} = 0.10 \pm 0.01$ and (d) $\bar{n}_{\alpha} = 20 \pm 0.13$.}
\label{Fig:temp_measurement}
\end{figure}

For evaluating the performance of the transport, it is important to have a reliable and efficient method for measuring the amount of motional excitation of the ion. While the dynamics of the ion shuttling is of classical nature, the precision which is required calls for energy measurements schemes which work quantum mechanically, i.e. down to the single phonon regime. Doppler recooling~\cite{Wesenberg2007} and ion loss rates have previously been used to measure extremely large energy transfers~\cite{HUBER2008}. Rabi oscillations on the motional sidebands of the Raman transition (see Fig.~\fref{temp_measurement}a) provide an accurate but time-consuming tool for reconstruction of the motional state~\cite{LEIBFRIED1996}. The measurement principle is based on the dependence of the Rabi frequency on the phonon number, which manifests itself in the Rabi oscillation signal:
\eq{rabi}{
P_{\uparrow,\Delta n}(\theta)=\tfrac{1}{2}\sum_{n=0}^{n_{\mathrm{max}}}p_n\left(1+\cos(M_{n,\Delta n}\theta)\right),
}
where $P_{\uparrow,\Delta n}(\theta)$ indicates the measured probability for finding the ion in the \ket{\uparrow} level after driving the stimulated Raman transition, \ket{\uparrow,n}$\rightarrow$\ket{\downarrow,n+\Delta n}, with a pulse area of $\theta$. The Rabi frequency is altered by the matrix element $M_{n,\Delta n}$~\cite{LEIBFRIED2003}. The phonon probability distribution $p_n$ is given by a convolution between a thermal and a coherent phonon distribution:
\eq{p_n}{
p_n = \sum_{m=n}^{N} \frac{\bar{n}_{th}^{m}}{(\bar{n}_{th}+1)^{m+1}} \cdot e^{-\bar{n}_{\alpha}} \frac{\bar{n}_{\alpha}^{(n-m)}}{(n-m)!},
}
where $\bar{n}_{th}$ is the thermal mean phonon number and $\bar{n}_{\alpha} = |\alpha|^2$ the mean phonon number arising from the coherent displacement, characterized by an amplitude $\alpha$, allowing for a distinction between the two distributions. The mean phonon numbers are extracted by performing a simultaneous Bayesian analysis of data sets pertaining to different transitions, $\Delta n$, for a given shuttling operation. Imperfect readout and preparation, as well as dephasing of the Rabi oscillations from other decoherence sources, are taken into account. Thus, estimations and valid confidence intervals for the mean phonon numbers, $\bar{n}_{\mathrm{th}}$ and $\bar{n}_{\alpha}$, are obtained for each data set.
For low excitation energies, probing is done on the transitions $\Delta n=-1,0,+1$ (see Fig.~\fref{temp_measurement}c), and in the high energy regime, $\Delta n=0,-1,-2$ is better suited (Fig.~\fref{temp_measurement}d).

As an optimized alternative, yielding the information about 100 times faster, we introduce a new energy metric, the \textit{pseudo energy}. It is based on $P_{\uparrow,\Delta n}$ for three different sideband transitions with fixed pulse areas:
\eq{pE}{
E_{\mathrm{p}} = 2 \cdot \left( P_{\uparrow,\mathrm{+1}}(3\pi) - P_{\uparrow,\mathrm{-1}}(3\pi) - P_{\uparrow,\mathrm{-2}}(2\pi) \right) + 7/2.
}
The indices refer to the blue ($\Delta n=+1$), red ($\Delta n=-1$), and second red ($\Delta n=-2$) motional sidebands, probed at the time it would take the carrier to reach the indicated pulse area. For phonon numbers below $\bar{n}_{\alpha} \lesssim 50$, this measure is positive and monotonically increases with the phonon number (see Fig.~\fref{temp_measurement}b) and is moreover robust against small fluctuations of the transition frequencies, laser beam intensities and readout imperfections. For mean phonon numbers $>$~50, the pseudo energy may still be used, but one must take care of the conversion ambiguity, visible in Fig.~\fref{temp_measurement}b.

\begin{figure}[ht]
\includegraphics[width=8.5cm]{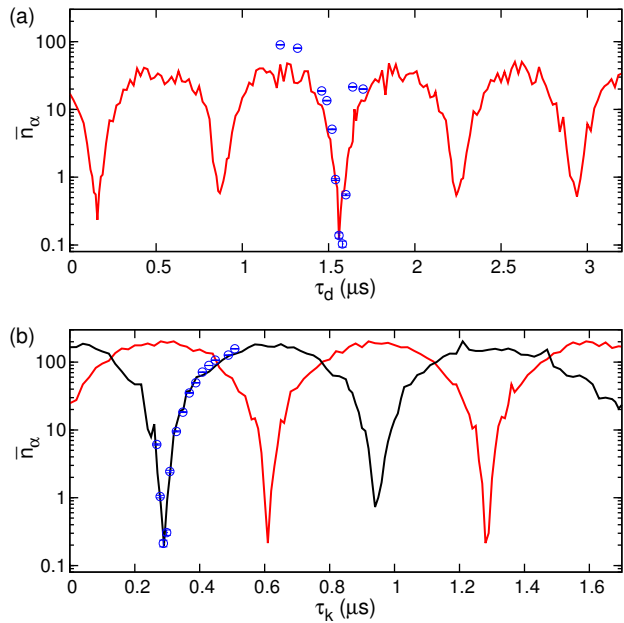}
\caption{(color online) a) Increase of motional energy for a pairwise neutral transport as a function of the dwell time between the two fast transports, with a transport duration of 11.2~\micro s in each direction. The solid lines show the phonon numbers obtained from the pseudo energy metric and the blue dots indicate phonon numbers obtained from the Rabi oscillation method. b) Energy increase for the self-neutral transport as a function of the kick delay time for when the stopping kick of +3.85~V (red) and of -3.68~V (black) is applied. The transport duration used here is 4~\micro s, with an adiabatic return transport of 180~\micro s. Slight drifts in the axial trap frequency affect the precise timing of the optimal kick delay such that a recalibration of the time axis was required.}
\label{Fig:pE_curves}
\end{figure}

We use the pseudo energy to investigate how the motional energy of the ion depends on a dwell time, $\tau_d$, between two fast transport operations, i.e. the pair-neutral scheme. The ion starts out in segment A (see Fig.~\fref{trap}a) and is then transported to segment B, where it remains for a time period that can be altered in steps of 20~ns (on top of the 400~ns update time), before being transported back to segment A where the readout takes place. The result is presented in Fig.~\fref{pE_curves}a. We observe a periodic dependence on the dwell time, with minima in intervals of $1/\nu_{\mathrm{ax}}$. A few points are evaluated more accurately using the sideband Rabi oscillation method, and for the dips we find minimum excitations of $\bar{n}_{\alpha} = 0.10 \pm 0.01$ phonons. When the dwell time instead is such that the momentum transfer from the first and the second transport add up, we find a maximum, where fits to the Rabi oscillations are consistent with a coherent state phonon distribution with up to $\bar{n}_{\alpha} \sim 100$ phonons. As expected, the energy transfer from the first shuttle can be coherently removed by the transport back, with a periodic dependence on the dwell time.

For the energy self-neutral shuttling, a sharp kick is used to stop the motion. This is realized by applying a voltage pulse on a nearby electrode segment, being active for only one update sample of the waveform generator (400~ns). The kick delay time, $\tau_k$, between the last voltage update of the transport and the stopping momentum kick is scanned. This effectively changes at which phase in the motional period the force is applied, thus either adding to or removing motional energy. The transport back to segment A is performed adiabatically, such that this part contributes only negligibly to the final excitation. The result is presented in Fig.~\fref{pE_curves}b, where both a positive and a negative voltage kick amplitude is shown, displaying a shift of half a cycle with respect to each other. The amplitude of the voltage kick is calibrated by minimizing the energy transfer at the kick delay time corresponding to the minimum. The self-neutral transports displays an excitation as low as the pair-neutral ones, but allows even faster shuttles. The shortest duration, for which transport with negligible excitation was achieved, was 3.6~\micro s, which corresponds to about 5 oscillation cycles, plus a delay time of $\tau_k=220$~ns.

For both types of transports, the results were obtained using ramps where the spatial location of the ion varies in a sin$^2$ shaped manner with respect to time. Other ramp shapes, such as a linear one, yield different results but are also able to realize energy-neutral transport, as considered previously~\cite{REICHLE2006,HUCUL2008}. We also applied both shuttling protocols to two-ion crystals, obtaining similarly low energy increases on both the common and stretch modes of oscillation. The results show that we obtain a high degree of control over the dynamics of trapped atomic particles. The placement accuracy in phase space is given by the minimum excitation of 0.1 phonons compared to the phase space volume occupied by the shuttling trajectory in units $\hbar$. From this we attain a relative accuracy on the order of 10$^{-8}$.

So far, we have shown that after a fast shuttle the motional ground state \ket{n=0} can be recovered.  This control is now extended to spin-motion entangled states: We investigate how quantum information can be stored, and transported, in a superposition of Fock states. For this we perform a spin echo experiment using pulses on the blue sideband Raman transition, where a pair-neutral transport is carried out in the second branch. Starting with the ion in state $\mket{\uparrow, n = 0}$, a $\pi/2$ pulse on the blue sideband will put the ion into the state $\mket{\uparrow, n=0} + e^{i\phi}\mket{\downarrow, n=1}$, a spin-motion entangled state with phase $\phi \equiv 0$. After the shuttling, a phase shift of this state will arise from changes to the energy levels of both the spin and the motional part of the wavefunction. If one can assure that the spin level splitting remains constant, any measured phase shift will reveal changes to the motional energy level spacing, making it a versatile and precise probe of the time-dependent trapping potential. In Fig.~\fref{bsb_spinecho} a and b, spin echo contrast curves are displayed, where the phase of the concluding $\pi/2$ pulse is varied. The data in Fig.~\fref{bsb_spinecho}a is obtained using the carrier transition with and without transport. After a magnetic field gradient compensation has been performed~\cite{Walther2011}, no phase shift between the two data sets is obtained, proving that the Zeeman splitting between the spin energy levels remains constant during the transport. The data in Fig.~\fref{bsb_spinecho}b is obtained using the blue sideband transition, and the phase shift visible in the transported case is thus acquired solely from the motional energy difference of the spin-motion entangled state.

\begin{figure}[ht]
\includegraphics[width=8.5cm]{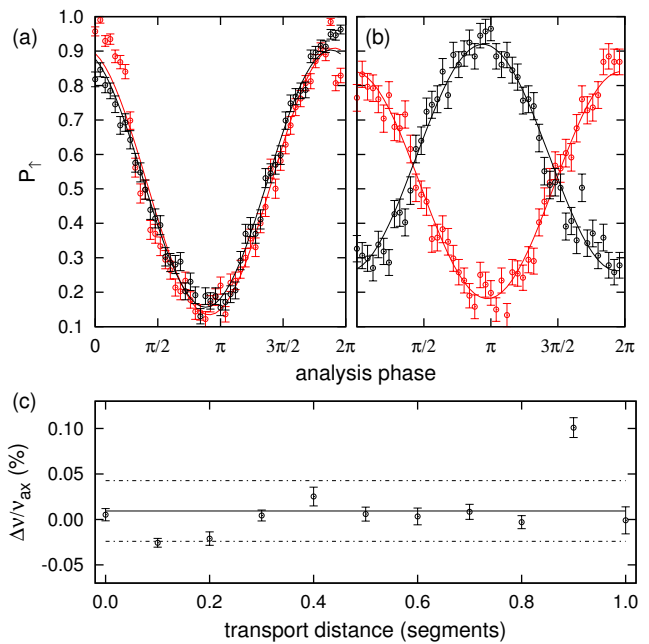}
\caption{(color online) Proof of phase coherence of the qubit during and after transport. a) Spin echo interference when the ion prepared in a superposition of spin states with $\left(\mket{\uparrow} + \mket{\downarrow}\right)\mket{n=0}$. No phase shift is observed when comparing the ion at rest (red) to a transport of one segment (280 \micro m) and back (black). b) When the ion is shuttled in a superposition of Fock states $\mket{\uparrow, n=0} + e^{i\phi}\mket{\downarrow, n=1}$, we observe a phase shift between the situation at rest (red) and with transport (black) indicating a change to the motional energy levels during transport. In c), the spin-motion entangled probe is used to obtain a distance-independent trap frequency, with precisions well below 1 kHz. The lines indicate a residual of less than one per mille of the trap frequency, where the dashed lines indicate the standard deviation of the distribution of points.}
\label{Fig:bsb_spinecho}
\end{figure}

Upon using the spin-motion entangled state as a sensitive probe for the phonon energy level splitting, we found that the trap frequency varied slightly with the distance between the segments, due to technical imperfections. This resulted in a spatial modulation of the trap frequency by at most about 80~kHz for an ion position right between the two segments, and near zero directly above the segment centers. A set of position dependent correction factors can be readily obtained, by scanning the dwell time at various distances and for varying voltages, such that the phase shift is zeroed at each location. After this procedure, the probing is repeated and we find that the trap frequency deviations are reduced by two orders of magnitude, down to the measurement precision of 390~Hz, using a dwell time of 100~\micro s. This corresponds to less than one per mille of the trap frequency, as shown in Fig.~\fref{bsb_spinecho}c. It is interesting to consider that the quantum information stored in a coherent superposition of two Fock states, even though being displaced by more than 100 phonons during the transport, can be fully retrieved.

We have demonstrated fast shuttling operations with small residual motional excitation, such that the ion ends the transport near the motional groundstate. We also showed that qubit information is preserved through the shuttling by transporting a spin-motion entangled state with negligible loss of coherence.  As a next step we intend to perform fast splitting operations of multi-ion crystals. Our long-term goal is to extend the transport scheme to parallel operations on multiple segments and combine it with laser-driven quantum logic gates.

The authors acknowledge Alex Wiens for earlier contributions and Heinz Lenk for development of electronics hardware. We acknowledge financial support by the IARPA SQIP project (MQCO framework), and by the European commission within the IP AQUTE and STREP Diamant, and the VW-Stiftung. A.W. acknowledges funding from the Swedish Research Council.

\bibliographystyle{apsrev4-1}
\bibliography{fast_transport_bib}

\end{document}